# DSSI FOR PILE SUPPORTED ASYMMETRICAL BUILDING: A REVIEW

Pallavi Badry[1] and Neelima Satyam D.[2]

[1]Resaerch Scholar,
[2]Assistant Professor,
Geotechnical Engineering Laboratory, Earthquake Engineering Research Centre,
International Institute of Information Technology.

**ABSTRACT**

*Several attempts have been carried out to understand the foundation failures during earthquake type of transient loading and it has been found that the interaction effect between soil and foundation is one of the important aspects need to be considered in the analysis and design. In this regard a literature survey has been carried out on frame structures supported on pile foundation which is in contact with more soil mass as compared to the other type of foundation system. In this research paper concept of dynamic soil–structure interaction is introduced, and the research methods were discussed. With the reference of the several documents in the field of soil structure interaction a document of present and past literature has been made with the including a main focus on interaction of pile supported frames. This study focuses on the complexity and excessive simplification of the model for foundation system and structures, and should be carried forward for its significance. The review is carried out including analytical, experimental and numerical approaches considered in the past study. The perusal of literature reveals that very few studies investigated on asymmetrical buildings supported on pile foundations. In this paper, an attempt is made to understand research carried out in pile soil structure interaction and research gap along with the scope of research has been identified to carry out the present research work.*

## 1. INTRODUCTION

In general analysis and design procedures of the foundations the type of footing depends up on the load induced by the superstructure, local site parameters and the earthquake zones. Generally the multi storied buildings constructed on weak strata are supported on pile foundation. The analysis of structure become more tedious if the interaction effect is included in the analysis, as in interaction analysis foundation system (Soil and foundation) and structure has to analyse with equal importance [70, 74]. The interaction problem can be analysed using direct and substructure method. In dynamic analysis, the total interaction response is the combination of the two parts namely kinematic and inertial interaction. When the seismic wave passes through a soil mass, it vibrates and displaces due to distortion and dilation of waves through the solid media, such vibration in the supporting soil mass is called as the kinematic interaction. Once the excitation wave passes through soil and enters in to the structure starts vibrating which exerts extra dynamic force to the soil mass, refers as the inertial interaction which depends up on the inertial forces produced by the structure [58,113].

## 1.1. Direct Method

In direct method the inertial and kinematic interaction is estimated by modelling soil and structure together. Kinematic interaction develops due to vibration of supporting soil mass and wave propagation through the soil whereas the inertial interaction causes once the vibration reaches to the base of structure and induces a vibration in the structure [58,113]. The response of the coupled system is calculated from the following governing equation given by Krammer (2003).

$$[M]\{Ü\} + [C]\{Ú\} + [K]\{U\} = -[M]\{Üg\} \tag{1}$$

Where, M, C and K are the mass, damping and stiffness matrices of the integrated system which includes the structure and foundation system.

Ü, Ú and U are the acceleration, velocity and displacement of the system and Üg is the ground motion acceleration.

## 1.2. Substructure or multistep method

In substructure method the interaction analysis has been divided in to several steps and the ultimate response of the system has been calculated by cumulating the response of the each step by using principal of superposition [57,71]. The response of the system due to kinematic interaction and inertial interaction can be calculated separately and then superimposed as per the theory of superstore method to get the combined system response [86]. Eq. 2 and 3 shows the Kinematic and inertial interaction separately.

$$[M_{soil}]\{Ü_{KI}\} + [C]\{Ú_{KI}\} + [K^*]\{U_{KI}\} = -[M_{soil}]\{Üg\} \tag{2}$$

$$[M]\{Ü_{II}\} + [C]\{Ú_{II}\} + [K^*]\{U_{II}\} = -[M_{structure}]\{Ü_{KI} + Üg\} \tag{3}$$

Where, $[M] = [M_{soil}] + [M_{structure}]$ and $[M_{structure}]$ is the mass matrix assuming soil is mass less, $[K^*]$ is stiffness of entire system, $[C]$ is damping matrix of entire system, Üg is input ground acceleration, $\{U_{II}\}$ displacement vector due to inertial interaction [62,65,105].

## 2. DAMAGES DURING EARTQUAKES

Several damages have been evidenced during earthquakes due to incomprehension of soil structure interaction effect in design of structure and foundation system [80,91,107]. Some of these are captured in this review paper. In the 1985 Mexico earthquake high rise building collapsed due to the partial bearing capacity failure of foundation soil. It has been reported that this earthquake was particularly destructive to the unbraced buildings founded on soft soils due to the increase in fundamental time period of soil from about 1.0 s to nearly 2.0 s induced due to the interaction phenomenon. In the 1995 Kobe earthquake (M=6.9) the interaction effect played a vital role in sudden increase of natural period where the collapse and overturning of Hanshin expressway is observed. In same earthquake Daikai station failed due to poor load transfer mechanisms due to interface effects [40]. In the 2010 Haiti earthquake (M=7), collapse of several buildings has observed because of deeper rotation failure due to movement of soils [37,41]. During 2001 Bhuj earthquake ($M_w$=7.6) caused extensive damage to life and property due to attenuation effect of the wave travelling through the soil layers with a high impedance contrast of the supporting soil layers [59].

# 3. FACTORS INFLUENCING SSI EFFECTS

Soil structure interaction is very complex phenomenon and its effect depends up on the many parameters including soil stratification, soil density, and wave propagation frequency. Few of these factors are discussed below.

## 3.1. Impedance contrast

Impedance contrast defined as the product of velocity and density of the material, thus varies the ground motion amplitude while travelling to the most heterogeneous soil media like soil [58].Seismic waves travels faster in hard rocks as compare to softer rocks and sediments [31,60]. As the waves passes from harder to softer media waves travels slower and in order to maintain the same earthquake energy attains the bigger amplitude [19,14,25].

## 3.2. Resonance

Resonance in earthquake phenomenon defined as the matching the magnitude of an excitation frequency (Frequency of earthquake wave) with the fundamental natural frequency of the system. Early attempts have been shown that the structural response against earthquake is different for fixed base analysis than the soil structure interaction analysis in frequency domain. [34,86,45].

## 3.3. Damping in Soil

In dynamic analysis when the excitation/seismic waves travel through the soil mass the energy of the wave is dissipated due to the scattering the waves in to the infinite domain. Thus the energy loss takes place in this phenomenon is called as the radiation damping. The energy of the input waves also can be used in deformations of the soil mass due to which the changes the soil material properties and referred as a material damping [28].Absorption of energy occurs due to inelastic properties of medium in which the particle of a medium do not react perfectly elastically with their neighbour and a part of the energy in the waves is lost instead of being transferred through medium, after each cycle [26, 58, 70].

## 3.4. Trapping of Waves

Impedance contrast between adjacent layers of soil mass is one of the important factors which cause the wave trapping in the soil mass. Kawase (1996) has brought in observation in the 1995 Hyogo-ken Nanbu earthquake which was the most destructive earthquake in Japan even though of moderate magnitude of 6 [8,26,30,58].

## 3.5. Lateral discontinuities

Lateral discontinuities can be explained as the softer material lies besides a more rigid one and vice versa. The damages were observed in the Bhatwari- Sonar village during the 1999 Chamoli earthquake due to the layer of debris dumped situated below the stiff soil [59].

The numerous research works have been carried out on soil structure interaction analysis founded on the different types of shallow and deep foundation configuration. It has been observed that much research gap is left with the attempts made on the interaction analysis of building founded on piles [20,21,22,23]. In this review paper an extensive publications have been featured to understand the status of the research in the area and the literature is summarized including the review domain as analytical, numerical, experiment and prototype observation.

# 4. LITERATURE REVIEW

## 4.1. Review on Experimental studies

Several researchers including Ward and Crawford (1964,1966) , Ivanovic (1995,2000), Yano (2000), Kusama (2003), Dunand (2004), Kontani (2004), Jayalekshmi (2009), Hokmabadi (2014) etc. carried out a research on a experimental full scale and scaled down the models to understand the soil structure interaction effects for the buildings with the different types of the foundation type [23,24]. The behaviour of the structure against the dynamic loading is best evaluated by the ambient and forced vibration tests [6]. These experimental models have been extended to the geotechnical applications to study the soil behaviour with the structure coupling under the dynamic load.

### 4.1.1. Ambient Vibration Tests

The ambient vibration tests describe the linear behaviour of the structures against the vibration produced by the sources including wind, microtremors, microseisms and various local random and periodic sources , since the vibration are small. The full scale test can be performed at a large and dense set of points by placing the seismometers in strategic locations throughout the building on both the directions. This test can be used extensively to identify and to monitor changes of system frequencies between small (ambient noise) and large (earthquake shaking) response amplitudes [27,36].

Crawford and Ward (1964) were among the first to show that the ambient vibrations test can be used to determine some of the lowest frequencies and modes of vibration of full scale structures [36]. This is achieved by carrying out an ambient vibration test on a nineteen story building against the random wind excitation and first three modes of vibrations have been studied. The results from the ambient test analysis for the building is tested against the analytical calculation and it has been concluded that the some of the theoretical and experimental values are in good agreement [24,54,72,83].

Several case studies including Kaprielian Hall building , Millikan Library building, 7-story reinforced concrete building in Van Nuys, California and many damaged building during earthquakes has been analysed by few researchers with the comparative study of modal frequencies of the building and concluded that reason for reduction of fundamental frequency of vibration may be the inducement of large strains following the strong shaking from earthquakes [94].The attempt has been made to understand the effect of SSI on damping cone model has been used. The model is validated against the ambient vibration test result for the building. The model has been extended for some realistic set of buildings and soil properties but it has been observed that the damping values so obtained cannot be extrapolated with the model data due to soil nonlinear behaviour and the large strains occurred in the soil under the ground shaking [24,47].

### 4.1.2. Forced Vibration Tests

The force vibration requires a large scale shaking to simulate the earthquake loading. The source always mounted at the top of the structure to get the more prominent excitation and modes of vibration that have amplitudes t the higher levels. The test is generally complete full scale tests. Full-scale vibration tests using large shaking machines have proved to be one of the most effective methods for determining the dynamic characteristics of structures needed for understanding the effects of strong earthquake motions [47,51].

Paul (1968) conducted the forced vibration tests on Millikan library building to understand its study state response. The test results concluded that the building responses similar as the fixed base analysis results and torsion at bottom and top of the building are negligibly [81, 82, 84].

In 1974 again the experiment has been conducted on the Millikan library building and building was forced to vibrate by a vibration generator located at the roof. The basement slab and the roof were measured for shaking in both NS and EW directions. The observations indicate that for NS excitations each floor remains essentially plane experiencing an almost uniform translation and uniform rotation about EW axis. Also results of experiment indicate that interaction of structure and soil has a marked effect on the response during forced vibration tests [82,84].

A full scale experimental study has been carried out for Hollywood storage building [104,105] and the results the researcher has concluded that the interaction problem is so complex to summarize the dependent parameter and to simplify it is needed to neglected few of them. In the research the rocking effect of the wave propagation is neglected and the response of the building is estimated depending up on the ground motion transfer function for different ground motions.

Few research has been carried out for estimating the significance of the SSI effects studied for both embedded and shallow foundation and revealed few facts like that the fundamental time period for due to SSI is a lengthen to 12% and 74% for the braced and unbraced structure, respectively [63,110].

Full scale vibration tests using random vibration methods has been used to predict the pile and soil response including interaction [2,4] . Experimental simulation techniques for characterizing dynamic soil pile interaction concluded that the new experimental techniques has good agreement with the traditional forced vibration test using shaker or vibrator.sts presented herein [50,54,83].

### 4.1.3. Shake Table Tests

The model test which carried out under the gravitational field of earth is known as shake table test. It is generally a prototype test in which the shake table is developed for a scale down model to simulate the earthquake and the response of the building is predicted. The test is well popular with some advantages such as well controlled large amplitude, multi-axis input motions, ease in performance and numerical simulation. The major work on in situ tests for SSI has been conducted to understand the behaviour of nuclear power plants at the time of dynamic loading. The Nuclear Power Engineering Corporation (NUPEC 1998) had conducted extensive experimental studies on the SSI of the nuclear power plants. The following are series of tests conducted to verify the seismic analysis codes [10,89,98]. The SSI for rigid structures has been studied by The parametric study for a dynamic loading including base mat size, dynamic soil stiffness, radiation damping and soil pressure distributions were investigated [115,38]. The similar study has been carried out to understand the uplift phenomenon including interaction effect for shallow raft without modelling a soil and found that the uplift pressure is developed in SSI analysis [11,9]. The shake table model is further extended for modelling a soil with silicon rubber. The dynamic soil structure interaction for different ground motions has been studied and study reveals that the contact ratio decreases with increasing input motions [9,38,39].

Yano (2000) performed a model tests on dynamic cross interaction tests to investigate dynamic cross-interactions of structures. Along with the above investigations some tests were also conducted to understand the dynamic cross interaction of adjacent buildings. Nuclear Power Engineering Corporation (NUPEC) of Japan has developed a shake table test model to demonstrate the sensitivity of structures of different heights to the frequency of the ground motion in a group. This study consists of both field and laboratory tests on scaled models. The field tests include forced vibration and earthquake observation. The laboratory tests involve vibration tests using an exciter and a shaking table test that applies simulated earthquake ground motions. The study has been concluded that the tendency that larger adjacent effects like displacements and rocking appear in the direction of parallel to the row of buildings [16].

Pitilakis (2007) designed a shaking table to confirm the ability of the numerical substructure technique to simulate the SSI phenomenon. The model includes SSI potential for embedded shallow foundation system supported by a dry bed of sand deposit poured in container. The experimental SSI system has subjected to a strong ground motion and the same geometry simulated numerically to validate the result of the complete soil foundation structure system with linear viscoelastic soil behaviour .Study concluded that the experimental and numerical responses in both frequency and in time domain were in good agreement [29,100]

The similar study has been extended for laminar box setup for a soil pile structure model with compacted soil domain as a elastic half space and the results are validated with the numerical model and found deviation in the test results are agreeable [10].

A single storey steel pile supported structure was tested for interaction performance. It has been noted that acceleration response of the pile cap was three times larger than that of the structural response. The pounding observed due to the development of a gap between soil and pile which led to the extraordinary large inertia forces at the top of the pile and pile failure occurred due to the cracks in the pile volume. The results concluded that the probable causes of pile damages is due to seismic pounding between the laterally compressed soil and the pile near the pile cap level [10,112].

Jayalekshmi (2009) studied the response behaviour of multi-storey structures with isolated or raft foundations, resting on a shallow soil stratum of homogeneous or layered soil, subjected to earthquake induced ground motions and to identify the structures in which the dynamic SSI effects are to be considered for a safe design. An experimental study has been carried out with the aid of a container box with structure embedded into the soil and accelerometers are placed in soil and building model for measuring the response to impact with impulse hammer [51].

The study concluded that the incorporation of soil flexibility in the analysis of low rise buildings is required for the realistic estimate of structural seismic response especially for single storey structures resting on very soft soil. Seismic base shear is found to decrease for the medium rise buildings, with raft foundation resting on soil as compared to fixed base due to longer natural period. It shows that the effect of SSI is advantageous to medium rise buildings on raft foundation. The investigations for low rise buildings with isolated footings resting on a shallow soil stratum on rock under transient loading reveal that the effect of SSI is to increase the seismic response of low rise buildings, up to three stories [51].

Madabhushi (2008) carried out a shake table test on a single pile in liquefiable soil strata. The results of this study provide information on performance criteria for seismic design of structures with pile foundations considering p-y relations for soil-pile interaction including the soil nonlinearity [48].Pile soil interaction has been studied against the sinusoidal shaking to the shake table and the response is calculated with the sensors at the different pile lengths using 1g shaking to the shear box including the liquefaction effect (Figure 1) [11].

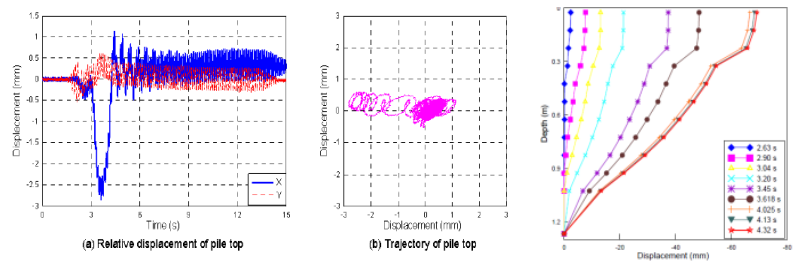

Figure 1. Response of pile with interaction effect [11]

Li, Hou and Liu (2012) carried out a shake table test with $1/15^{th}$ scale down model on 3 by 3 pile group as a foundation type on a single symmetric building and the two adjacent building. The study concluded that an adjacent structures experiences more serious damage than that in single one. The SSI effect has some influence on the soil frequency and the damping ratio of the SSSI system, but has little influence on the frequency and the characteristics of the vibration modes of the SSI system [115].

Hokmabadi (2014) carried out a series of shake table tests for a mid rise symmetrical building supported by a shallow and deep foundation system in a homogenous soil condition. The response of the system is estimated for different ground motions and compared with the fixed base condition. The study concluded that the SSI effect is significantly high in shallow foundation than the deep foundation due to resistance soil and pile resistant [116].

### 4.1.4. Centrifugal Tests

The test model which performs under higher gravitational field is known as centrifuge tests. The widespread technique of centrifuge modelling applied to geotechnical structures helps to investigate complex engineering problems. The static soil structure interaction problem for deep foundation system can well be modelled using this techniques but dynamic loads generated by shocks and earthquakes are difficult to model in centrifuge. The behaviour of small-scale centrifuge models is representative of the behaviour of a full scale structure, named prototype. By increasing the mass forces, the centrifuge replicates the intensity of the prototype's stress and strain fields in scaled earth structure models (Phillips, 1869). Many researchers outlined the principle aspects of centrifuge modelling in geotechnical sciences (e.g., Schofield 1980; Craig 1985; Corté 1989a, b; Taylor 1995; Garnier (2001, 2002), Hajialilue (2007). The similitude relations govern the scaling relationship between the model and the prototype. Garnier et al. (2007) and Ellis and Aslam (2009) compiled the principles, aspects, and uses of similitude relations in a state of the art. Briançon and Simon (2012) present the results of in situ experimental studies on rigid pile reinforced embankments (Okyay et. al 2014).

Zelikson (1983) studied a scaled model of existing power station supported on its raft foundation and compared the response of the same model founded on anchor piles. Gosh (2007) conducted tests on different types of soil stratifications supporting a rigid containment structure. The results indicate that accelerations transmitted to structure base are dependent on stiffness degradation in supporting soil [117]. Chang (2006) studied the effectiveness of the commonly adopted retrofit strategy of adding a shear wall to a reinforced concrete frame through centrifugal modelling. Data analyses of centrifuge tests indicate that frame wall systems have highly asymmetric hysteric loops due to asymmetry of lateral force resisting system (Shear wall is in asymmetrical position on one end of building) [12].

Hajialilue (2007) established a test set up to find out the soil pile interaction effect in installation operations including the different installation procedures for the pile [99].The study reveals that the interaction effect and pile response varies with the different installations procedures (Figure 2).

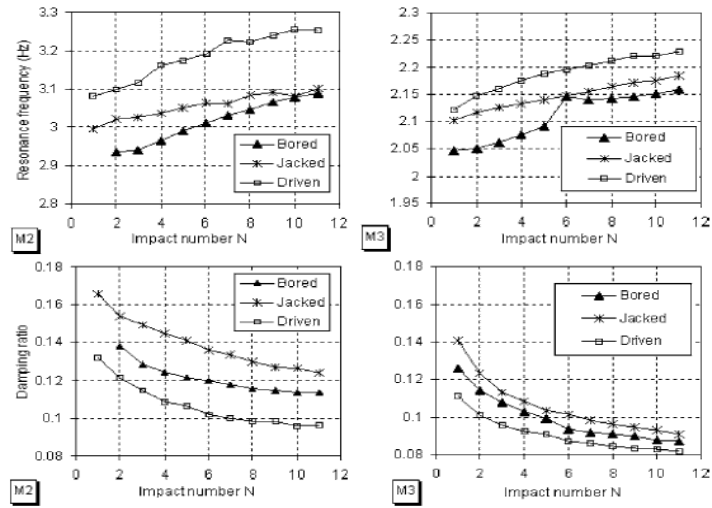

Figure2. Interaction effect for Bored, Jacked and Driven piles [99]

Ellis and Aslam (2009) performed centrifuge tests with both 6 and 16 vertical rigid piles under an embankment. In every test, the acceleration level ranged from 10-60 times g to simulate the different configurations of the prototype. They concluded that increasing the earth-platform's height decreases the differential settlement at the surface of the embankment [1].

Kutter (2013) used a large geotechnical centrifuges to model realistic soil, foundation and super structures. Two models were tested a moment frame with shear wall building and soil pile bridge deck interaction. New techniques imposed in the centrifugal modelling give results that match with the existing systems [16].

### 4.2. Review of Analytical studies

Madhubhushi (2008) has given a good amount of theoretical procedures to understand the dynamic behaviour of the pile including the interaction effects [42]. The many traditional techniques are available to analyse the pile response for both under static and dynamic loads including interaction. Some of the methods are briefed as follows.

#### 4.2.1. Winkler Approach

The Winkler approach (1867), also called the sub grade reaction theory, is the oldest method to predict pile deflections and bending moments. In this method the soil is represented by the series of the elastic springs with a spring constant ($K_h$) equal to the soil subgrade reaction can be calculated theoretically and experimentally.

Poulos and Davis (1980) and Prakash and Sharma (1990) carried out the work on the response of the single pile under static and dynamic loading conditions and developed a standard charts which can be used to determine pile response in terms of deflections, slopes, and moments. Though the Civil Engineering and Urban planning:An International Journal(CiVEJ)Vol.1, No.2/3, December 2014work has been appreciated for the approach but it is not popular in practice because of its theoretical shortcomings and limitations in estimation of Kh which is not a unique property of the soil, but depends intrinsically on pile characteristics and the magnitude of deflection. The approach is imperial which includes the independent spring behaviour thus the response so calculated is not the true and realistic [7].

### 4.2.2. *p - y* Method

The p-y approach to analysis of response of laterally loaded piles is essentially a modification of the basic Winkler model. In this approach p is the soil pressure per unit length of pile and y is the pile deflection. McClelland and Focht (1956) solved the beam bending equation and obtained a nonlinear load versus deflection curves to model the soil. This approach is known as the p-y method of analysis. The various commercial software such as COM624 and LPILE Plus3.0 adopted this techniques in Finite Element Program to estimate the nonlinear response of the pile [7]. In this method the soil is represented by a series of nonlinear p‐y curves that vary with depth and soil type. Reese (1977) has developed a number of empirical curves for typical soil types based on the results of field measurements on fully instrumented piles.The pile response against the loading in terms of deflections, rotations and bending moments are calculated by solving the beam bending equation using finite difference or finite element numerical techniques [87,52].

### 4.2.3. Elastic Continuum Approach

Poulos (1971) analyzed the behaviour of laterally loaded piles with the elastic continuum approach where the approach valid for the homogeneous elastic soil only [65,87]. The soil pile model has been developed by assuming pile as a thin rectangular vertical strip divided into elements and each element is acted upon by uniform horizontal stresses which are related to the element displacements through the integral solution of Mindlin's formulation [87].

Soil pressures over each element are obtained by solving the differential equation of equilibrium of a beam element on a continuous soil with the Finite Difference Method (FDM) which gives the subsequent solution for the displacements.Novak (1974) extended the approach including the effect of soil structure interaction. Several attempts have been carried out with continuum approach including constant stiffness and equivalent viscous damping for single [65]. Madhav and Sarma (1982) developed a program to study the behaviour of overhang pile embedded in homogeneous soil mass subjected to both axial and lateral loads. The study reveals that the load displacement behaviour was found to be dependent on magnitude of axial load and also on pile and soil parameters including height of overhang, relative stiffness of pile and soil, undrained shear strength [63].

### 4.2.4. Finite Element Method

The finite element method is a numerical approach based on elastic continuum theory that can be used to model pile soil pile interaction. The complexity in handling a large scale model is one of the advantages of the numerical technique [32]. Finite element techniques have been used to analyze complicated loading conditions in which the soil is modelled as a continuum. Pile displacements and stresses are evaluated by solving the classic beam bending equation using one of the standard numerical methods such as Galerkin, Collocation, or Rayleigh Ritz. In finite element methods various type of element are used to represent the different structural components like soil, pile, frames etc. [51]. Time dependent results can be obtained by modelling the various types of structure and geometry. The special interface or contact techniques need to adopt to model a interaction problem [32].The method can be used with a variety of soil stress strain relationships, and is suitable to analyze group response of the piles.

### 4.3. Review of Numerical Studies

Almost all structures are in contact with soil, thus it is essentially important to understand the effect of supporting strata on the superstructure. The SSI problem is very complex due to combined effect of several factors like soil nonlinearity, interface behaviour, soil impedance, wave propagation effect, structural inertia, Viscous boundary etc, hence it is highly impossible to estimate the response of the system nearest to the reality. Thus to understand the SSI problem clearly, and model the system close to the reality the versatile tool is need to be adopted. The finite element method is one of the numerical techniques which give the flexibility to model the complexity which is essential for SSI problem.
In following sections the numerical different approaches and techniques have been disused to model the SSI problem with more realistic way.

#### 4.3.1. Numerical Modelling of buildings

The widespread availability of powerful computers has brought about a great change in the computational aspect recently. Numerical methods are widely scoped than that of analytical methods. The methods are so versatile that it captures the all possible complexity to model many conditions with a high degree of precision, realism, including nonlinear stress strain behaviour, non homogeneous material conditions, and changes in geometry etc. A numerical model has been developed to estimate the response of the ground along with the pile foundation and concluded that pile experiences a mobility and large displacements in the direction of wave propagation [14].

Lehmann and Antes (2001) used the FEM BEM coupling model to investigate the structure soil structure phenomenon by giving a vertical load in the soil between the two structures. The study reveals that found that this hybrid numerical model is suitable for analyzing the structure soil structure interaction problems. Emphasis was an application of FEM BEM coupling methods to SSI problems when more than one structure is present. But no such clarity is given in this study [32,60].

The study has been carried out to understand the seismic behaviour of tall buildings including the non linearity in soil and pile behaviour during strong earthquakes where a 20 storey building is examined as a typical structure supported on a pile foundation for different configurations like rigid base, linear soil pile system and nonlinear soil pile system. The effects of pile foundation displacements on the behaviour of tall building are investigated, and compared with the behaviour of buildings supported on shallow foundation [98]. The study has been extended for large embedded foundations in dams and tall buildings and observed that rocking and translation plays a major role in changing the response of system and it has been concluded that the rigid foundation doesn't show any movements [96].

A combined Finite element based SSI model with consistent infinitesimal Finite Element Cell Method is used for modelling the soil region extending to infinity (far-field), and the standard Finite Element for the finite region (near-field) and the structure. This method is implemented numerically to analyse the structural response subjected to the harmonic and transient forces which included the interaction effect. The model accuracy has been checked with various soil regions homogeneous, layered and top layer resting on a rigid bed rock. In order to decrease the computation time and achieve the solution of large scale problems, the model is parallelized. As a result of this parallel solution, significant time is saved for large scale problems [94,95,98].

A parallel SSI model using a finite element based computer code has been proposed which contains finite and infinite elements for bounded and unbounded soil regions respectively. A sub structure method is applied to the soil structure system and the domain is represented by separate sub structures and interfaces. To check the accuracy of the model the results are compared with boundary element method .The study emphasized on validation of parallelized computer code by studying the already existing problems for which analytical solution is given. No significant comments about SSI phenomena have been given [87,100].

The research has been carried out to understand the response of the massive foundation when it is subjected to the dynamic loading by FEM BEM coupling approach. This approach has been proved to be very effective to model interaction phenomenon and study also reveals the Soil structure interaction is much governs the response in case of massive and heavy foundation system[81,74].Gazetas (2004) studied the dynamic analysis of the rocking response of SSI problem by finite element discretization using ABAQUS. Soil behaviour is represented with the elastoplastic Mohr-Coulomb model. It has been concluded that the initiation of uplifting and the mobilization of bearing capacity failure can be quite beneficial for the superstructure, under certain conditions related with the fundamental period of the structure and characteristics of ground shaking. In this study a nonlinear behaviour in the dynamic SSI analysis, sliding at the soil foundation interaction is not considered, but in reality this also incorporates nonlinear behaviour on soil foundation interaction [75,78].

Maheswari (2004) developed a numerical model for single pile and 2 x 2 symmetric pile groups and soil nonlinearity taken in to account by HISS material model and the soil and pile interaction is achieved by Kelvin element. The study reveals the facts that for a harmonic excitation, the soil nonlinearity increases the pile head and structural responses at low frequencies [5].

A new approach has been developed to carry out the analysis for separation or gap creation induced at the pile head due to the inertia forces from the structure using time domain Winkler soil model (Hyperbolic soil constitutive model). The study observed that correction is necessary while dealing with separation of pile from soil domain during excitation. Effect of separation on the response depends very much on the level of nonlinearity, since accordingly various states i.e. separation, yielding and delinking are determined. Due to separation, there is increase in displacement while force is decreased which shows that stiffness of soil is decreased. Separation decreases the dynamic stiffness of the soil-pile system. As the level of nonlinearity increases, separation becomes more intense heading towards yielding and delinking and thus increasing the gap [5,17].

Kham (2006) studied the site city interaction using 2 D boundary element method subjected to vertically incident wave. To investigate such phenomena, called site city interaction (SCI) herein, two simplified site city configurations are considered, one building configuration with same building and same spacing between them and other configuration in which different buildings with different spacing between them. These 2D boundary -element method models are subjected to a vertically incident plane SH waves. It has been observed that building density and city configuration play a major role in energy distribution inside the city [26,31,34]. A soil pile interaction using Winkler's nonlinear model numerically and the response of the pile has been compared with the 3-D FEM model and centrifuge model [41].The study concluded that once the behaviour enters in plastic condition the p-y model shows large deviation from the experimental results whereas 2D FEM approach has good agreement with the experimental results [45].

Wegner (2009) studied structural response of a two way asymmetrical multi-storey building model subjected to bi directional harmonic and earthquake loadings. The response in terms of vertical and horizontal displacements and rotation of the roof has been studied obtained using the scaled boundary finite-element method. The program Dynamic Soil Structure Interaction Analysis (DSSIA-3D, Wegner, 2009) which takes into account the soil structure interaction effects is applied to study two way asymmetrical buildings. These results are compared with those of symmetrical buildings. The study has drawn the observation that lower and medium size asymmetrical buildings (up to 15 storey) buildings incurred the most impact from the earthquakes. The mass effect is not a major influential factor for tall buildings. It has been observed that asymmetrical building coupled with the two way asymmetrical earthquake loadings amplified the damages to the structure compared with symmetrical buildings [108].

Padron et al. (2009) carried out a study to understand the SSI for the adjacent building. To understand the behaviour the coupled BEM-FEM model has been generated and response id studied. The research reveals the fact that SSI effects on group of structures with similar dynamic characteristics are important and also system response can be either amplified or attenuated according to the distance between adjacent buildings. Mao (2009) modelled a real time building in Fujian province of China and the effect SSI has been cofigurated. The peak response of absolute acceleration, story drift, moments at beam ends, as well as inner force of columns and shear walls are analyzed under two orthogonal horizontal directions seismic excitations. The SSI effect with nonlinearity on seismic response of high-rise building is summarized and the rationality of reduction factor for soil-structure interaction calculation specified. Along with this several researchers carried out a SSI analysis accordance with the guidelines given in Chinese seismic code however, the seismic response of structural member may be amplified in some cases, but most of the studies found the response in the safer limits it [92,102,110,107].Pitilakis (2010) proposed an equivalent linear substructure approximation of the soil–foundation–structure interaction system. A numerical code MISS3D has been developed to perform soil–foundation–structure interaction analyses in the three-dimensional linear elastic or viscoelastic domain, based on the substructure method. MISS3D is extended in order to model the nonlinear soil behaviour through an equivalent linear approach, resulting in a numerical tool named MISS3D-EqL. The proposed approximation is established theoretically and then validated against centrifuge tests [72,54,55].

Maheshwari (2011) carried out a soil pile interaction in liquefiable soil medium. Three dimensional soil pile-structure systems modelled with a finite element code developed in MATLAB. A Kelvin element (spring and dashpots) has been used to model the radiation boundary conditions including the soil nonlinearity effect by work hardening plastic cap model. The study concluded that the bending moment owing to liquefaction is drastically increased at the transition zone between the liquefying and non-liquefying soil medium. Christoph Knellwort, Herve Peton and Lyesse Laloui (2011) discusses several is issues pertaining to heat exchanger piles. The study explains the geotechnical numerical analysis method based on the load transfer approach that assesses the main effects of temperature change on pile behaviour.

Mohmoud Ghazavi, Omid Tavasoli (2012) presented numerical analysis of pile driving for tapered piles. Y. Xiao, L Chen(2012) discussed steel H shaped piles which are widely used in bridge foundation. The study has found the experimental results from monotonically loaded static tests on model steel H pile to pile cap connections including the interaction effect , in which the piles were subjected to tensile loading or horizontal loading with the bending in the strong or weak bending directions of the H pile. The tests indicate that H pile footing connections were effective in transferring vertical and lateral loads. The study also show that FEM analyses can capture the load and deformation relationship and load carrying capacity of the steel H pile to pile-cap connections satisfactorily. Yaru Lv, Hanlong Liu, Xuanming Ding and Ganpolang Kong F(2012) investigated the behaviour of X-section cast-in-situ concrete piles to estimate the settlement due to interaction effect. It has been found that the XCC can significantly increase ground-bearing capacity.

Kampitsis et al (2013), developed a 3-D New-Beam approach based on the Boundary Element Method (BEM) to describe the phenomenon of soil pile structure interaction. The study explains the soil stratification with the soil interface as a Kelvin Voigt element including the soil nonlinearity implemented by formulations of a hybrid spring configuration consisting of a nonlinear (p–y) spring connected in series to an elastic spring damper model. The approach is validated with the existing numerical tool and found that the approach is efficient and accurate [48,66,76,85].

Varun et.al (2013) developed a macro element for soil structure interaction analyses of piles in liquefiable soils and results found to be good agreeable range when compared with the physical model [105]. Hokmabadi (2014) carried out a 3-D nonlinear numerical analysis on the friction pile including the soil pile interaction for different ground motions and concluded that the base shear which will be the stability measure for the superstructure will be more in friction piles as compared to the shallow foundations and frames supported with the frictions piles achieves a better life safety limit as compared to the other type of foundations (Figure 3) [2,116].

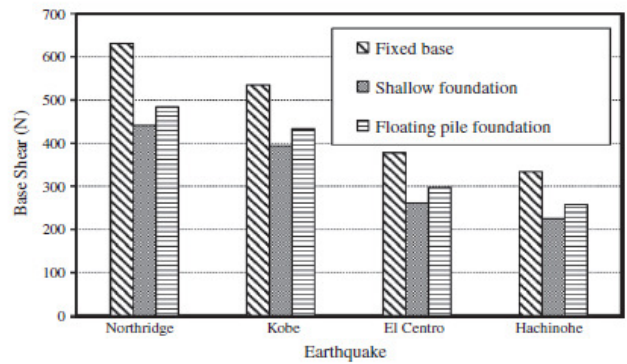

Figure 3. Base shear of the model structure obtained from 3D numerical analysis for: fixed-base structure; structure supported by shallow foundation; and structure supported by floating (frictional) pile foundation, [116].

In all the above studies it has been have considered the buildings as shear type buildings and the soil as homogeneous, linear, elastic medium. But the dynamic structure soil structure interaction of nearby buildings has not been studied by considering fully framed structure with heterogeneous soil, material nonlinearity, nonlinearity in the super structure.

**4.3.2 Numerical methods for efficient computation**

There are several numerical techniques including Finite Element Method, Boundary Element Method, hybrid are available to work out the problem of soil structure interaction.

Bathe (1986) introduces the efficient use of Finite Element Method for complex geometry. Now FEM is an efficient common computing method widely used in civil engineering, discretizes a continuum into a series of elements with limited sizes to compute for the mechanics of continuum. A new numerical method developed after FEM only discretizes the boundary of the definition domain. It is different from the discretization of total continuum and uses functions satisfying the governing equation to approximate boundary conditions [68,114]. The BEM is more advantageous compared to FEM because it requires only a surface discretization and satisfies automatically the radiation condition without any need for using special complicated non-reflecting boundaries as required by FEM by Wang S (1992).

In order to achieve the computational efficiency in modelling the high volume problems like soil structure interaction the two versatile methods like BEM and FEM are coupled, owing to respective disadvantages of FEM and BEM. The coupling method of FEM and BEM was developed in the field of SSSI in 1990s. The methods prove to be very efficient to simulate superstructures, foundations and near field soils whereas BEM is used for far field soil [67,117]. At present there are a large number of available commercial finite element programs including ANSYS, ABAQUS, MSCMARC which have friendly interface and powerful nonlinear solver.

## 5. CONCLUSIONS

### 5.1. Research Gap

From the detailed literature review on soil structure interaction for pile supported structures, the following are the broad occlusions have been drawn

1. SSI study has been carried out by few the pile supported mid rise building (15-20 storied structure) supported on the homogeneous soil strata by few researchers considering linear and non linear soil behavior both has independently addressed.
2. SSI consideration for linear and non linear soil behavior for both static and dynamic load conditions is attempted for both superstructure and foundation with some limitations like effect of water table in the analysis. But detailed analysis considering the effect of water table is not been addressed.
3. Not much work has been carried out on SSI analysis for asymmetrical structures has supported by pile foundation.
4. Few researchers has presented study on the SSI analysis for bridges with deep foundation System like pile only considering the soil behavior as a linear and homogenous, the study is not extended for the soil nonlinearity and heterogeneity.
5. A very few studies have been addressed on slender and Asymmetrical buildings with pile foundation. It has been observed that asymmetrical building with shallow foundation is been presented by few researchers but no attempt is made for deep foundation system.
6. Considering the present literature on pile soil interaction the research on complete three dimensional models representing the soil-pile-frame structure interaction system with nonlinear soil model is not carried out so far.

### 5.2. Scope of the work

As per the state-of art literature review on the Soil Structure Interactions analysis, it has been noted that among the available techniques Numerical modelling can best simulate the soil structure interaction phenomenon close to the reality. But limitation of this technique is computational time required for the interaction analysis. The following points which need to address for carrying out the research on Soil Structure Interaction analysis for asymmetrical tall structure supported by pile foundation are as follows.

— Soil Structure Interaction analysis for the asymmetrical building supported with the pile foundation system.
— Soil Structure Interaction analysis for the tall symmetrical building supported by the pile foundation system.
— Soil Structure Interaction analysis considering the torsion in foundation system for the symmetrical and asymmetrical buildings.
— Soil Structure Interaction analysis considering the soil heterogeneity needs to carry out considering the effect of pore water.

Civil Engineering and Urban planning:An International Journal(CiVEJ)Vol.1, No.2/3, December 2014

The scope of the present study is to study Soil Structure Interaction effect for the pile supported asymmetrical buildings in the stratified soil. The study focuses to present a new numerical approach to optimize the computational time to get the results of the integrated SSI model with the direct approach.

Civil Engineering and Urban planning:An International Journal(CiVEJ)Vol.1, No.2/3, December 2014[20] Chore, H. S. and Ingle, R.K. (2008), ″Interactive Analysis of Building Frame Supported on Pile Group Using Simplified F.E. model″ ,Journal of Structural Engineering, SERC, Vol. 34(6), pp: 460-464.

[21] Chore, H. S., Ingle, R. K. and Sawant, V. A. (2009), ″Building Frame-Pile Foundation-Soil Interaction Analysis″, Interaction and Multi scale Mechanics, Vol. 2, pp: 397-411.

[22] Chore, H. S., Ingle, R. K. and Sawant, V. A. (2010), ″Building Frame-Pile Foundation-Soil Interaction Analysis: A Parametric Study″ , Interaction and Multi Scale Mechanics, Vol. 3, pp: 55-79.

[23] Crouse, C.B, Liang, G.C and Martin G.R, (1984), "Experimental study of soil-structure interaction at an accelerographs station", Bulletin of the Seismological Society of America, Vol. 74, pp: 95-113.

[24] Crouse, C.B., Hushmand, B., Luco, J.E., and Wong, H.L. (1990), "Foundation impedance functions: Theory versus Experiment", Journal of Geotechnical Engineering, ASCE, Vol.116, pp: 432-449.

[25] Crouse,C.B, McGuire,J., (2001), "Energy Dissipation in Soil-Structure Interaction", Earthquake Spectra, Vol. 17, pp: 235-259.

[26] D. G. Lin and Z. Y. Feng (2006), "A numerical study of piled raft foundations", Journal of the Chinese Institute of Engineers, vol. 29, pp. 1091–1097.

[27] D. Givoli (1992), "Numerical methods for problems in infinite domains",  Elsevier Science Publishers B.V.,Amsterdam.

[28] D. Pitilakis, M. Dietz, D.M. Wood, D. Clouteau, A.Modaressi (2008), "Numerical simulation of dynamic soil-structure interaction in shaking  table testing" , Soil Dynamics and  Earthquake Engineering, Vol. 28, pp: 453-467.

[29] Day, S.M. (1978), "Seismic response of embedded foundations," , Proceedings in ASCE Convention, Chicago, IL, October, Vol. 34 (50).

[30] De Barros. C.P. and Luco, J.E (1995), "Identification of foundation impedance functions and soil properties from vibration tests of the Hualien containment model,", Soil Dynamics and Earthquake Engineering, Vo. 14, pp: 229-248 153.

[31] Deepa B. S., Nandakumar C.G. (2012), ″Seismic Soil Structure Interaction Studies on Multistorey Frame", International Journal of Applied Engineering Research and Development, Vol.2, pp: 45-58.

[32] Desai, C. S., and Kuppusamy, T. (1980), ″Application of a Numerical Procedure for Laterally Loaded Structures″, Numerical Methods in Offshore Piling, London, Vol. 4, pp:93- 99.

[33] Desai,C.S. Kuppusamy, T. and Allameddine A. R (1981), ″Pile Cap- Pile Group- Soil Interaction″, Journal of Structural Division, ASCE, Vol.107,pp: 817-834.

[34] E. M. Comodromos, M. C. Papadopoulou, and I. K. Rentzeperis (2009), "Pile foundation analysis and design using experimental data and 3-D numerical analysis",  Computers and Geotechnics, Vol. 36, pp: 819–836.

[35] Ivonac, Trifunac, Todavoska (2000),  "Ambient Vibration tests of structures-A review", ISET Journal of earthquake Technology, Vol. 37, pp:165-197.

[36] Rovithis, E., Kirtas, E., and Pitilakis, K. (2009), "Experimental p-y loops for estimating seismic soil-pile interaction" , Bulletin of Earthquake Engineering, Vol.7, pp:719-736.

[37] Iguchi, M., Akino, K., and Noguchi, K. (1987), "Model Tests on Interaction of Reactor Building  and Soil", Proceedings in 9th International Conference on Structural Mechanics in Reactor Technology

[38] Gazetas, G. (1991). "Formulas and charts for impedances of surface and embedded foundations," Journal of Geotechnical Engineering, Vol.117, pp:1363-1381.

[39] Gerolymos, N., Escoffier, S., Gazetas, G., and Garnier, J. (2009), "Numerical modeling of centrifuge cyclic lateral pile load experiments", Proceedings of the ICE - Bridge Engineering, Vol. 162,pp: 35-47.

[40] H. R. Tabatabaiefar and A. Massumi (2010), "A simplified method to determine seismic responses of reinforced concrete moment resisting building frames under influence of soil-structure interaction", Soil Dynamics and Earthquake Engineering, Vol.30, pp. 1259–1267.

[41] H. Tahghighi and K. Konagai (2007), "Numerical analysis of nonlinear soil-pile group interaction under lateral loads",  Soil Dynamics and Earthquake Engineering, Vol. 27,pp: 463–474.

[42] Hetenyi, M. (1946),  ″Beams on Elastic Foundation″,  The University of Michigan Press, Ann Arbor, Michigan.

[43] Horvath, J. S. (1984), ″Laterally Loaded Deep Foundations: Analysis and Performance" American Society for Testing and Materials, Vol.112, pp: 112-121.

[44] Hushmand, B. (1983), "Experimental Studies of Dynamic Response of Foundations ", PhD thesis, California Institute of Technology, Pasadena, CA.
60